\documentclass[journal,transmag]{IEEEtran}

\usepackage{cite}
\usepackage{graphicx}
\usepackage{amssymb,amsmath}

\begin{document}

\title{Iron and gold thin films: first-principles study}


\author{
	\IEEEauthorblockN{
Justyna Rychły-Gruszecka, 
Hubert Głowiński,
Justyn Snarski-Adamski,
Piotr Kuświk,
Mirosław Werwiński}	
	\IEEEauthorblockA{Institute of Molecular Physics, Polish Academy of Sciences, Smoluchowskiego 17, 60-179 Poznań, Poland}
	}

\IEEEtitleabstractindextext{%
\begin{abstract}
Using density functional theory, we carried out systematic calculations for a series of ultrathin iron layers with thicknesses ranging from one atomic monolayer to eleven monolayers (up to about 1.5~nm). We considered three cases: (1) iron layers both on a gold substrate and coated with gold, (2) iron layers on a gold substrate but without coverage and (3) free-standing iron layers adjacent to a vacuum. For our models we chose initial bcc Fe(001) surfaces and fcc Au(001) substrates. Based on the calculations, we determined the details of the geometry and magnetic properties of the systems. We calculate lattice parameters, magnetic moments, Curie temperatures and magnetocrystalline anisotropy energies. From the thickness dependence we determined the volume and surface contributions to the magnetic anisotropy constant. The further analysis allowed us to determine the thickness ranges of the occurrence of perpendicular magnetic anisotropy, as well as the effect of thickness and the presence of a substrate and cap layer on the direction of the magnetization easy axis.
\end{abstract}

}

\maketitle

\pagestyle{empty}
\thispagestyle{empty}

\IEEEpeerreviewmaketitle


One of the most important properties of a magnetic thin film is its magnetocrystalline anisotropy energy (MAE) since it is one of the key factors determining the orientation of magnetization at remanence.
Sufficiently strong anisotropy with the axis of easy magnetization perpendicular to the surface of the layers allows perpendicular magnetization of the sample in remanence, which, for example, is required in magnetic tunnel junctions to maximize resistance output.
Thin film systems with perpendicular anisotropy, due to the possibility of obtaining magnetic textures with perpendicularly magnetized very narrow domains, are also important for data storage applications.
In this context, layered systems are particularly interesting because of their ability to tune effective material parameters, such as magnetic anisotropy.
Among such systems, iron-based layered heterostructures are of significant interest.

First-principles calculations of iron layers on gold have been going on since the 1980s~\cite{li_monolayer_1988}.
It was predicted, for example, that in Fe/Au multilayers above the thickness of three atomic monolayers of iron, there is a change in the sign of the MAE interpreted as a change in the direction of magnetization of the layer from perpendicular (perpendicular magnetic anisotropy - PMA) to lying in the plane of the layer~\cite{szunyogh_magnetic_1995}.
The calculations showed a significant increase in the magnetic moment values of several atomic layers at both the Fe surface and Fe/Au interfaces.
Earlier calculations, however sophisticated, usually failed to reproduce accurate magnetic results due to the potential shape approximations used, as well as the most commonly assumed bulk values of lattice parameters and the lack of optimization of geometries of layers and heterostructures.

\begin{figure}
\centering
\includegraphics[clip,width=\columnwidth]{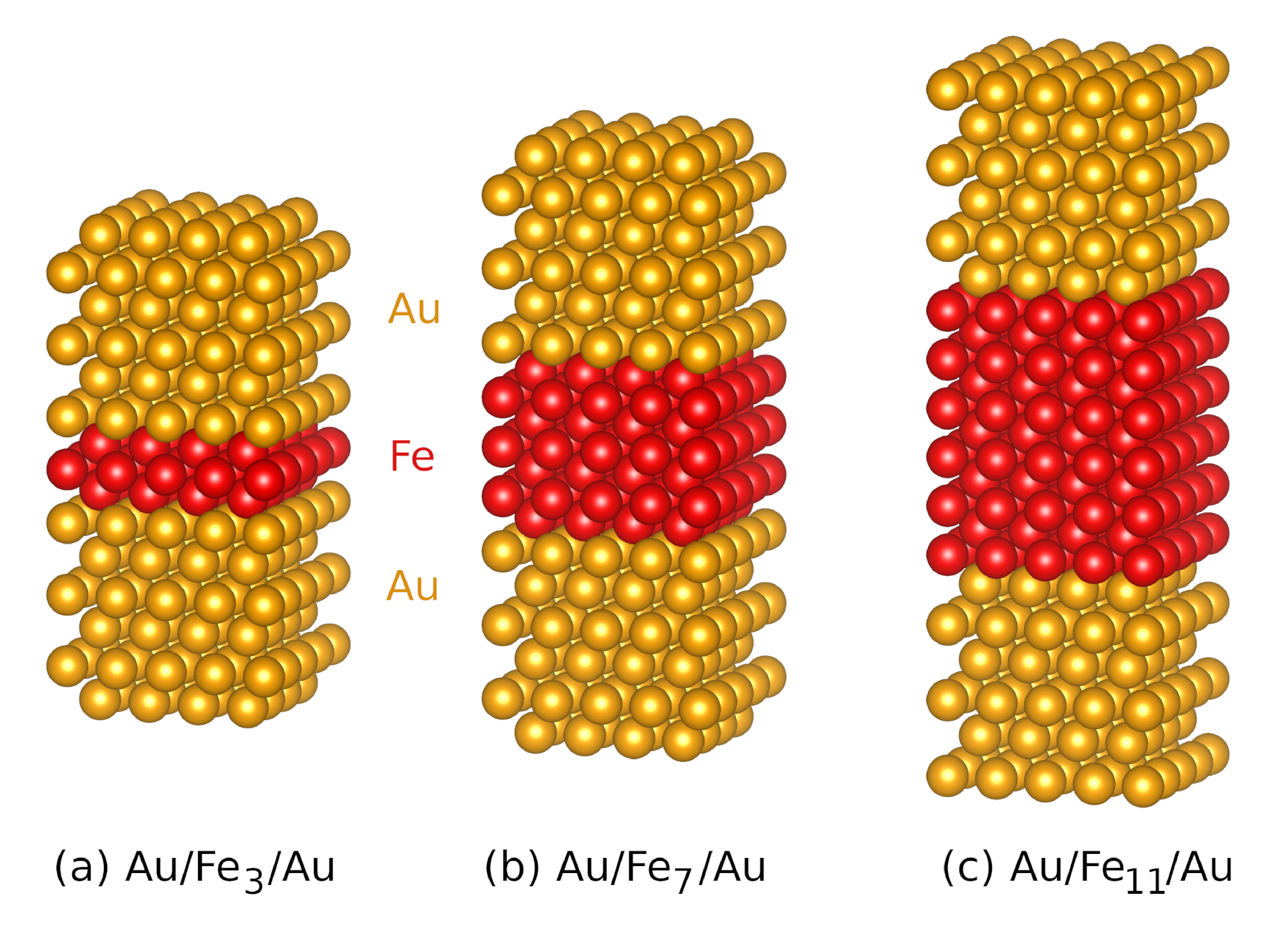}
\caption{
Selected structural models of Au$_{11}$Fe$_n$ multilayers. 
In each case, several unit cells are shown in the layer plain. 
In contrast, only one unit cell is shown in the direction perpendicular to the layer plain.
}
\label{fig:au11fex_vac_novac_comparison} 
\end{figure}

%
In this work we performed first-principles calculations for different thicknesses of iron layers deposited on fcc-gold substrates.
We have considered a range of iron layers from single atomic monolayer up to eleven monolayers. 
In all the heterostructures considered, the thickness of the gold layers equals eleven atomic monolayers. 
We consider three types of structures.
The first type is sandwich-type structures -- for which layers of iron and gold are alternated with each other, see Fig.~\ref{fig:au11fex_vac_novac_comparison}.
The second type is iron layers with two different surfaces - surrounded by a gold substrate on one side and a vacuum on the other. 
The third type is free-standing iron layers adjacent to a vacuum.

We performed density functional theory (DFT) calculations using the full-potential local orbital (FPLO) code~\cite{koepernik_full-potential_1999} in the FPLO18.00-52 version. 
Using Hellmann–Feynman forces, we optimized the lattice parameter $a$ and the Wyckoff positions $z$ which allowed us to determine the lattice parameter $c$.

\begin{figure}
\centering
\includegraphics[clip,width=0.8\columnwidth]{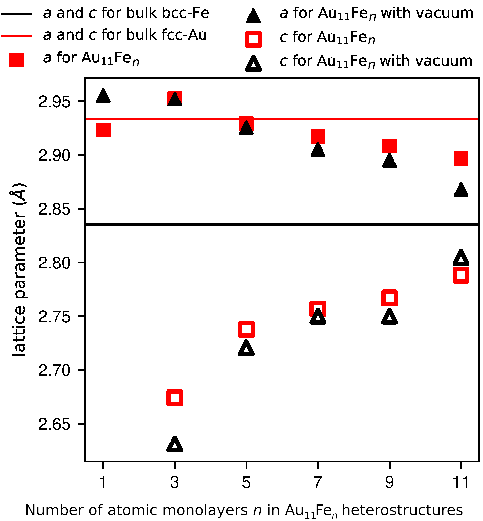}
\caption{
Lattice parameters calculated for Fe layers in Au$_{11}$Fe$_n$ multilayers and for Fe layers in Au$_{11}$Fe$_n$ heterostructure with vacuum (uncovered Fe surface). 
Calculations were performed with the FPLO18 code using the PBE functional.
}
\label{fig:a_c_vs_n} 
\end{figure}

%
%
\begin{figure}
\centering
\includegraphics[clip,width=0.8\columnwidth]{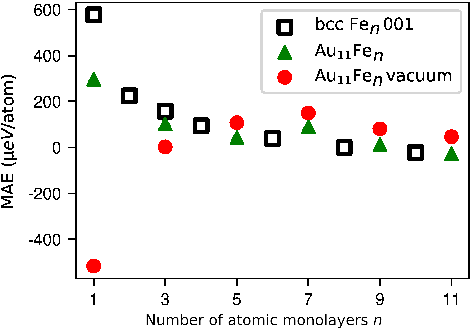}
\caption{
Magnetocrystalline anisotropy energy (MAE) as a function of Fe layer thickness (expressed as the number of atomic monolayers \textit{n})
for Fe layers in Au$_{11}$Fe$_n$ multilayers and 
for Fe layers in Au$_{11}$Fe$_n$ heterostructure with vacuum (uncovered Fe surface).
Results for free-standing Fe layers are included as a reference.
Calculations were performed with the FPLO18 code using the PBE functional.
}
\label{fig:MAE} 
\end{figure}

Figure~\ref{fig:a_c_vs_n} shows the dependence of lattice parameters on Fe layer thickness. 
Assuming perfect epitaxial growth of Fe layers on Au substrate in the model, the heterostructure model is described by a single unit cell with a lattice parameter $a$ common to Fe and Au layers. 
In the case of gold layers, the presented lattice parameter $a$ defines an fcc Au unit cell in a body centered tetragonal (bct) representation.
Due to the influence of the Au substrate, the obtained $a$ ($c$) values are higher (lower) than the values for the bulk Fe structure. 
The highest (lowest) values of obtained $a$ ($c$) occur for the thinnest one-atom-thick bcc-Fe layer. 
(In addition, the figure shows the values of lattice parameters $a$ for bulk fcc Au and bcc Fe.)
The thinner the Fe layer, the more its structure deviates from cubic toward a flattened tetragonal cell with a lattice parameter ratio $c/a$ of about 0.9.
As is known from previous studies, such a relatively strong deformation due to the influence of the substrate and overlay layer can have a large impact on the value and direction of magnetic anisotropy.

%
Figure~\ref{fig:MAE} shows the dependence of magnetocrystalline anisotropy energy (MAE) on the thickness of the Fe layer for both Au/Fe multilayers, as well as for Fe layers on Au substrate without a cap Au layer.
Multilayers tend to magnetize out-of-plane for thin Fe layers. 
However, for thicker Fe layers they magnetize in-plane. 
This is similar behavior to that found for free-standing Fe layers.
However, this relationship is reversed for Fe layers on Au substrate surrounded by vacuum. 
The above result points to the presence of a cap layer as another important factor affecting the direction and value of magnetic anisotropy.

%
%
\begin{figure}[]
\centering
\includegraphics[clip, width=0.9\columnwidth]{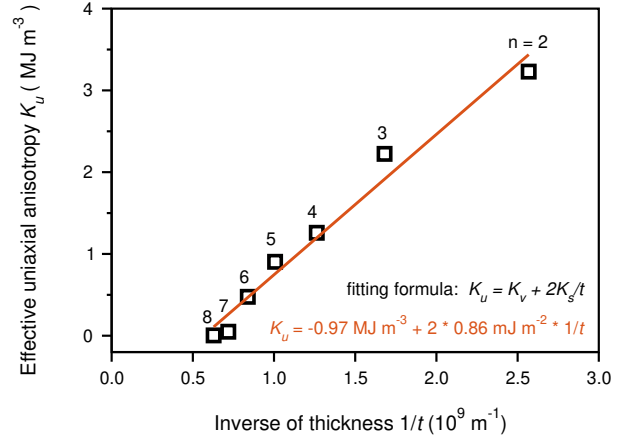}
\caption{\label{fig:ks}
Effective uniaxial anisotropy ($K_u$) as a function of inverse thickness (1/$t$) for free-standing Fe layers with thickness of two to eight monolayers (bct phase). 
Fit allows determination of surface magnetic anisotropy ($K_s$).
}
\end{figure}
Figure~\ref{fig:ks} shows the effective uniaxial anisotropy ($K_u$) of thin Fe films as a function of the inverse of thickness (1/$t$). 
Fitting with the equation $K_u = K_v + 2K_s/t$~\cite{kyuno_first-principles_1996} allows us to determine the surface magnetic anisotropy ($K_s$), which is equal to 0.86~mJ\,m$^{-2}$ and is in good agreement with experimental results.
It is known that the value of $K_s$ strongly depends on the type of interface.
Heinrich~et~al.~\cite{heinrich_magnetic_1991} observed the strongest surface anisotropy $K_s$ for the Fe(001)/vacuum interface (0.96~mJ\,m$^{-2}$ (0.96~ergs/cm$^2$)), followed by the Fe/Ag (0.81~mJ\,m$^{-2}$) and the Fe/Au (0.47~mJ\,m$^{-2}$) interfaces, all at room temperature. 
At lower temperatures (77~K), the latter two results are slightly higher (no low $T$ data for the Fe/vacuum interface).

\section*{Acknowledgement}
We acknowledge the financial support of the National Science Center Poland under the decision DEC-2018/30/E/ST3/00267 (SONATA-BIS 8). 
Part of the computations were performed on resources provided by the Poznan Supercomputing and Networking Center (PSNC).

\bibliographystyle{IEEEtran}
\bibliography{au_fe}

\end{document}